# Chemical-scissor-mediated structural editing of layered transition metal carbides

Haoming Ding[1,2,3], Youbing Li[1,3], Mian Li[1,3], Ke Chen[1,3], Kun Liang[1,3], Guoxin Chen[3], Jun Lu[4], Justinas Palisaitis[4], Per O. Å. Persson[4], Per Eklund[4], Lars Hultman[4], Shiyu Du[1,2,3], Zhifang Chai[1,2,3], Yury Gogotsi[5]\*, Qing Huang[1,3]\*

[1]Engineering Laboratory of Advanced Energy Materials, Ningbo Institute of Materials Technology and Engineering, Chinese Academy of Sciences; Ningbo, Zhejiang 315201, China.

[2]University of Chinese Academy of Sciences; 19 A Yuquan Rd, Shijingshan District, Beijing 100049, China.

[3]Qianwan Institute of CNiTECH; Ningbo, Zhejiang 315336, China.

[4]Thin Film Physics Division, Department of Physics, Chemistry, and Biology (IFM), Linköping University; SE-581 83 Linköping, Sweden.

[5]Department of Materials Science and Engineering and A. J. Drexel Nanomaterials Institute, Drexel University; Philadelphia, PA 19104, USA.

\*Corresponding author. Email: huangqing@nimte.ac.cn; yg36@drexel.edu

**Abstract:** Intercalation of non-van der Waals (*vdW*) layered materials can produce new 2D and 3D materials with unique properties, but it is difficult to achieve. Here, we describe a structural editing protocol for 3D non-*vdW* layered ternary carbides and nitrides (MAX phases) and their 2D vdW derivatives (MXenes). Gap-opening and species-intercalating stages were mediated by chemical scissors and guest intercalants, creating a large family of layered materials with unconventional elements and structures in MAX phases, as well as MXenes with versatile termination species. Removal of surface terminations by metal scissors and stitching of carbide layers by metal atoms leads to a reverse transformation from MXenes to MAX phases, and metal-intercalated 2D carbides. This scissor-mediated structural editing may enable structural and chemical tailoring of other layered ceramics.

Intercalated materials are predominantly produced by introducing non-native species into the van der Waals (*vdW*) gaps of inherently layered *vdW* materials such as graphite, hexagonal boron nitride (hBN), and transition metal dichalcogenides (*1, 2*). Guest-host interactions alter the electronic structure and enable property tailoring for energy storage, catalysis, electronic, optical, and magnetic properties (*3-8*). $M_{n+1}AX_n$ or 'MAX phases' for short, are a large family of ternary layered compounds which have weak metallic bonds between M and A atoms and covalent bonds within the MX layers (*9, 10*). Here, M denotes an early transition element, A is a main group element, X is nitrogen and/or carbon, and n is typically 1-3. The strong non-*vdW* bonding in MAX phases requires chemical etching of A elements to obtain 2D MXenes (*11-13*). The resultant *vdW* gaps in MXenes provide space for intercalating various guest species. For instance, anions such as $F^-$, $O^{2-}$, $OH^-$ and $Cl^-$, spontaneously coordinate with exposed M atoms of MXenes as termination species, T, as described in formula $M_{n+1}X_nT_x$ (*11, 14*). Intercalation of cations, cationic surfactants, and organic molecules in *vdW* gaps expands the interlayer spacing of MXenes and facilitates their delamination into monolayers, finding applications in energy storage, printed electronics, electromagnetic interference shielding, and many other applications (*15-19*).

Recently, we have reported a Lewis acidic molten salt (LAMS) etching protocol that is capable of both etching and substituting weakly bonded interlayer atoms in MAX phases (*14*). A series of MAX phases containing unusual late transition metals and MXenes with pure halogen terminations were synthesized and explored for catalysis, ferromagnetism, and electrochemical energy storage(*20-23*). In previous studies, LAMSs acted as an etchant to draw A-site atoms out of MAX phases and simultaneously provide ligands, such as -Cl, -Br, and -I, for MXenes and metal atoms, such as Zn, Cu, Fe, for MAX phases. However, the feasibility of LAMS etching depends on their thermophysical properties (solubility, melting point, and boiling point) and chemical properties (redox potential and activity of cations, coordination of anions), and only a few of them satisfy the dual role of both etchant and intercalant. For example, MXenes with terminations of -O, -S, -Se, -Te, and -NH were only realized by an anion-exchange reaction with brominated MXenes(*24*). The direct use of oxide or chalcogenide with strong covalent bonds to supply -O and chalcogen terminations would be a daunting task due to their high melting temperature and low solubility, which significantly limits the structural editing capability of LAMS etching. Here, we introduce a chemical-scissor-mediated intercalation chemistry for structural editing of non-*vdW* MAX phases and *vdW* MXenes (Fig. 1A). The etching and intercalating steps are separately regulated, and a mutual structure transition between MAX phases and MXenes is achieved. The constituent elements of MAX phases and terminating groups of MXenes are greatly extended, as indicated in the periodic table of elements (Fig. 1B). Structural editing by alternate use of LAMS and metal scissors leads to delamination of MXenes directly in molten salts, and also guides the discovery of a series of 2D metal-intercalated layered carbides.

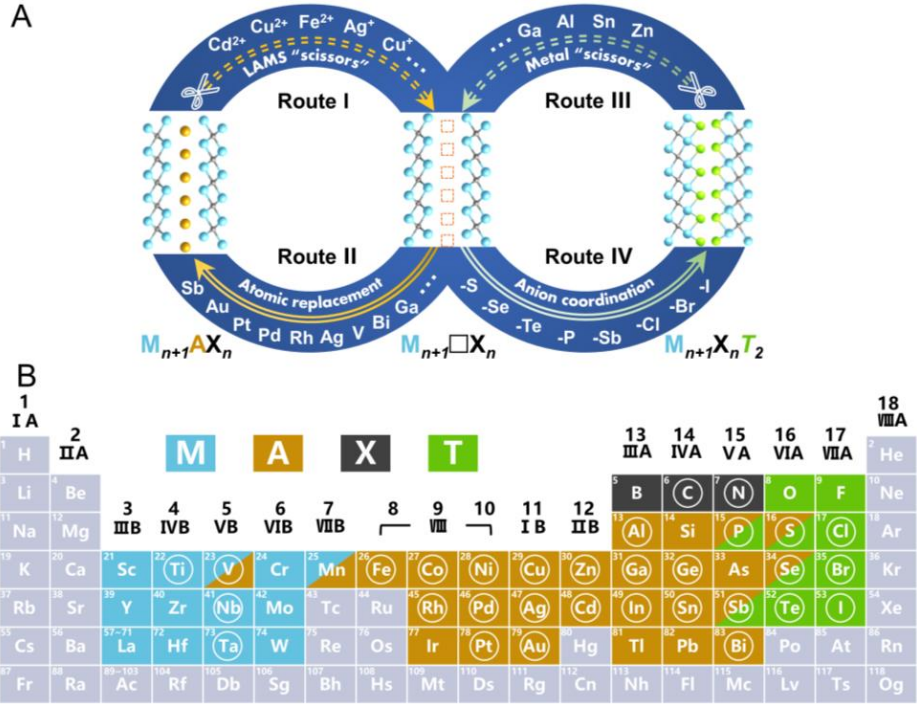

**Fig. 1. Structural editing protocol of MAX and MXene mediated by the chemical scissors.** (**A**) Schematic illustration of structural editing of MAX phases and MXenes through chemical scissor mediated intercalation protocol. (**B**) Periodic table of elements showing elements involved in the formation of MAX phases and MXenes. Light blue: M elements, brown: A elements, black: X elements, green: ligand elements, circled: elements studied in the present work.

Figure 1a illustrates the structural editing protocol of MAX phases and MXenes through chemical-scissor-mediated intercalation. The underlying chemical intercalation mechanism is the following: (1) the opening of non-*vdW* gaps in MAX phases by LAMS scissors due to redox potential difference between Lewis acidic cations and A elements (**Route I**); (2) spontaneous inter-diffusion of metal atoms into gaps to form new MAX phases to lower the system chemical energy (**Route II**); (3) remove of surface terminations of multilayer MXenes by metal scissors and the opening of *vdW* gaps (**Route III**); (4) the coordination of anions with oxidized early transition metal in *vdW* gaps to form terminated MXenes (**Route IV**). Figure 1b shows the periodic table of elements, highlighting elements that form MAX phases and MXene. The circled elements are investigated in the present work, and their recipes are listed in Table S1.

In more detail, cations (*e.g.*, $Cu^{2+}$) in a LAMS scissor (*e.g.*, $CuCl_2$) have a strong electron withdrawing capability to oxidize interlayer atoms that are weakly bonded in MAX phases (*e.g.*, $Cu^{2+}+2e^-=Cu$; $M_{n+1}AlX_n=M_{n+1}\square X_n+Al^{3+}+3e^-$) (**Route I**). As soon as interlayer atom vacancies $V_A$ or gaps $M_{n+1}\square X_n$ are available, pre-dissolved guest metal atom A′ (*e.g.*, Pt) in molten salts can diffuse into gaps in $M_{n+1}\square X_n$ and occupy $V_A$ positions to form new MAX phases (*e.g.*, $M_{n+1}\square X_n+V_A+Pt=M_{n+1}PtX_n$,) in an isomorphous replacement reaction (**Route II**). Otherwise, if exposed M atoms in $M_{n+1}\square X_n$ further lose electrons and reach high oxidation states (*e.g.*, $M_{n+1}\square X_n= M_{n+1}\square X_n^{2+}+2e^-$; $Cu^{2+}+2e^-=Cu$), they accept the non-bonding electron pair from Lewis base anions T⁻ (*e.g.*, $Cl^-$, $Br^-$ and $I^-$) in molten salts and form stable coordination structures described by the formula $M_{n+1}\square X_n^{2+}+2T^-=M_{n+1}X_nT_2$ (**Route IV**).

The *vdW* interaction between terminated carbide layers and strain created due to intercalation of ions into atomically thin galleries between MXene sheets account for the accordion morphology of multilayer MXenes. The exchange of ligands by coordination-competing reaction offers an option to modify the terminations and thus tune material properties. Moreover, these ligands in $M_{n+1}X_nT_x$ can be knocked out by scissors of reductive metals with a low electron affinity which reduce M atoms in $M_{n+1}\square X_n$ back to a lower oxidation state (*e.g.*, $Ga=Ga^{3+}+3e^-$; $M_{n+1}X_nT_2+2e^-=M_{n+1}\square X_n+2T^-$) (**Route III**). The elimination of ligands and breaking of interlayer *vdW* interaction in multilayer MXenes is accelerated by the evaporation of gaseous products (*e.g.*, $Ga^{3+}+3Cl^-=GaCl_3\uparrow$) with low boiling temperature (*e.g.*, ≈ 474K for $GaCl_3$), which activates the diffusion of metal atoms into gaps $M_{n+1}\square X_n$ and the restoration of crystal structure of MAX phase (*e.g.*, $M_{n+1}\square X_n+Ga=M_{n+1}GaX_n$) (**Route II**). Similarly, the M atoms on the surface of $M_{n+1}\square X_n$ derived from MXene can again become coordination centers for ligands after oxidation (**Route IV**). Besides main group elements (such as Ga, In, Sn, Ge, Bi, and Sb), which have been commonly used to synthesize MAX phases, unconventional transition metal A elements (Fe, Co, Ni, Cu, Zn, Pt, Au, Pd, Ag, Cd, Rh, and V) can also be intercalated to form MAX phases. Furthermore, the termination species -P and -Sb (group 15 elements), in addition to the known halogen (-Cl, -Br, -I) and chalcogen (-S, -Se, -Te) terminations, have been realized for MXenes.

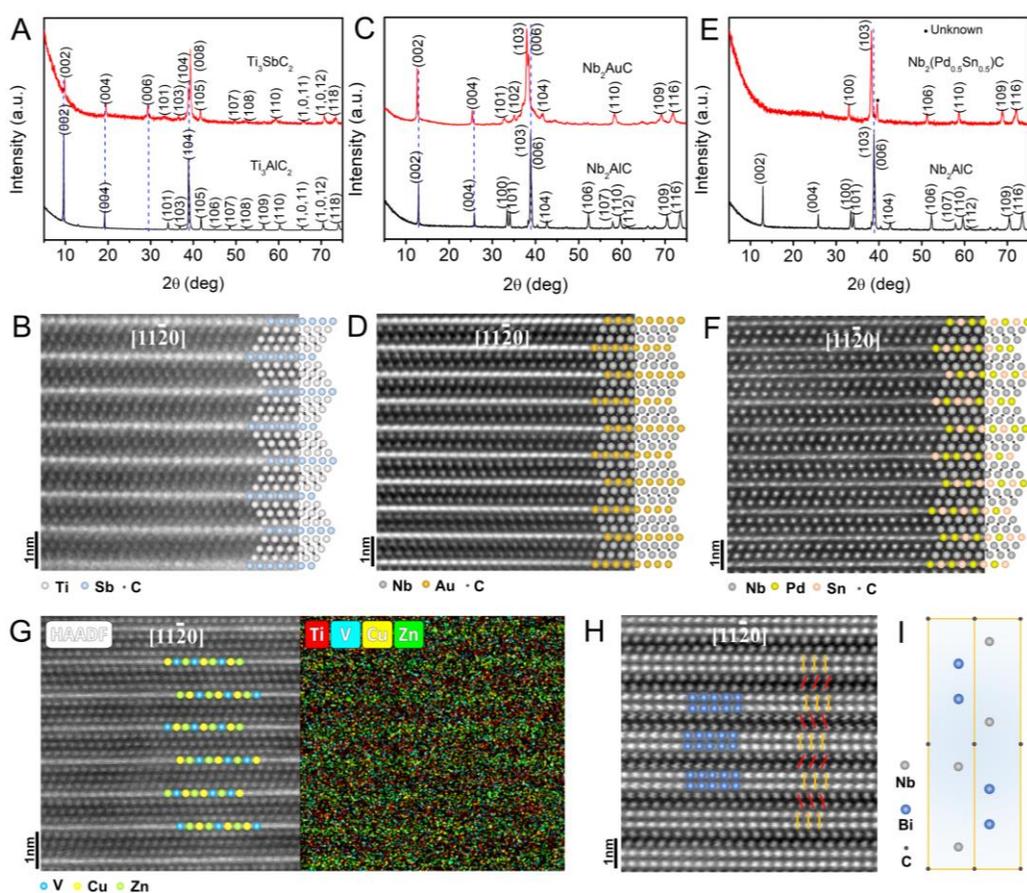

**Fig. 2. Transformation of a MAX phase to another MAX phase.** (**A**) XRD patterns of $Ti_3SbC_2$ and its parent phase $Ti_3AlC_2$. (**B**) High-resolution (HR)-STEM image of $Ti_3SbC_2$ along the [11$\bar{2}$0] zone axis with its atomic structural model. (**C**) XRD patterns of $Nb_2AuC$ and its parent phase $Nb_2AlC$. (**D**) STEM image of $Nb_2AuC$. (**E**) XRD patterns of $Nb_2(Pd_{0.5}Sn_{0.5})C$ and its parent phase $Nb_2AlC$ (**F**) STEM image of $Nb_2(Pd_{0.5}Sn_{0.5})C$. (**G**) STEM image of $Ti_3(V_{0.4}Cu_{0.3}Zn_{0.3})CN$ and its corresponding atomic resolution EDS mapping showing the elemental distribution. (**H**) STEM

image of $Nb_2Bi_2C$ and (**I**) its unit cell.

**Isomorphous replacement of MAX phases**

Using a $CuCl_2$ LAMS scissor, Al atoms in series of $M_{n+1}AlX_n$ phases (*e.g.,* $Ti_3AlC_2$, $Ti_3AlCN$, $Nb_2AlC$, $Ta_2AlC$) were etched out (**Route I**), and then the $V_{Al}$ vacancies in gaps $M_{n+1}\square X_n$ are re-occupied by ambient main-group elements in molten salts (**Route II**) to form new $M_{n+1}A'X_n$ phases (A'= Ga, In, Sn, and Ge) (figs. S1 to S5). Note that the low melting points of these main-group elements facilitate intercalation kinetics of atom diffusion into $M_{n+1}\square X_n$ gaps. Furthermore, the LAMS scissor must preferentially etch the A atoms (*e.g.,* Al) out of MAX phases but avoid the oxidation of intercalating metals (*e.g.,* Bi, Ge, Sn, In, and Ga) that can be determined from the Gibbs free energy calculation (fig. S6)(*25*). According to the above intercalation chemistry, Sb is successfully incorporated, creating unknown MAX phases, such as $Ti_3SbC_2$, $Ti_3SbCN$, $Nb_2SbC$ and $Ti_3(Sb_{0.5}Sn_{0.5})CN$ (figs. S7 to S12). In the X-ray diffraction (XRD) pattern of $Ti_3SbC_2$, the 000*l* peaks shifted towards higher Bragg angles compared to the $Ti_3AlC_2$ precursor (Fig. 2A), indicating a shrinkage of *c* lattice parameter from 18.578 Å for $Ti_3AlC_2$ to 18.443 Å for $Ti_3SbC_2$ (fig. S13). Scanning transmission electron microscopy (STEM) image of $Ti_3SbC_2$ shows the atomic positions which well match that of the MAX phase along the [11$\bar{2}$0] direction (Fig. 2B). $Ti_3C_2$ sheets are separated by monoatomic layers of Sb atoms, which are much brighter than Ti atoms due to the higher atomic number of Sb. Lattice-resolved STEM energy dispersive spectroscopy (EDS) and scanning electron microscopy (SEM)-EDS further showed the absence of Al in final $Ti_3SbC_2$, indicating the complete substitution of Al by Sb through the Lewis acid mediated intercalation chemistry (fig. S14).

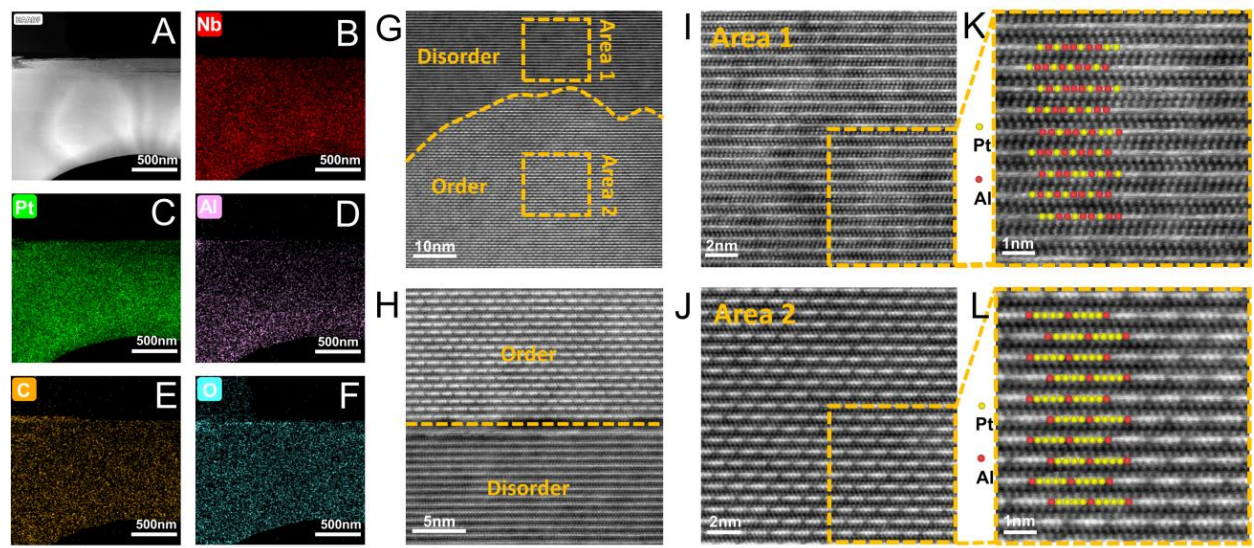

**Fig. 3. Transition from disorder to order of A-layer atoms.** (**A**) High-angle annular dark field (HADDF) image of $Nb_2(Pt_{0.6}Al_{0.4})C$ and corresponding EDS maps of Nb (**B**), Pt (**C**), Al (**D**), C (**E**), and O (**F**) elements. (**G-H**) STEM images of different areas of $Nb_2(Pt_{0.6}Al_{0.4})C$, showing the distinct boundaries between the disordered and ordered areas. (**I-J**) Partial enlarged images of Area 1 and Area 2 in (**G**), showing the disordered and ordered atomic arrangements, respectively. (**k-l**) Partial enlarged images of (**I**) and (**J**), respectively, with the atomic models that show the atomic arrangement in A-layers.

Noble metal elements are seldomly considered for synthesis of MAX phases due to their inert chemical reactivity and high melting temperature(*26*). Formation of an alloy with a low eutectic point can accelerate the diffusion of refractory noble metal atoms. For instance, the LAMS scissor $CdCl_2$ can open the non-*vdW* gap of $Nb_2AlC$, and itself be reduced into Cd metal. The Cd-Au alloy formed in molten salt has a eutectic point of 629 °C, which is much lower than the melting point of gold (1064 °C) and also below the synthesis temperature (700 °C). This facilitates intercalation of gold atoms into the gap of $Nb_2\square C$ and formation of fully-substituted $Nb_2AuC$ (Figs. 2, C and D), as well as partially-substituted $Nb_2(Au_{0.5}Al_{0.5})C$ (figs. S15 to S18). Similarly, late transition metals in the fourth period, such as Fe, Co, Ni, Cu, and Zn, can form alloys with Cd, which have low eutectic points. MAX phases including $Nb_2FeC$, $Nb_2CoC$, $Nb_2NiC$, $Nb_2CuC$ and $Nb_2ZnC$ were synthesized here (figs. S19 to S27).

When the metal reduced from LAMSs cannot sufficiently lower the eutectic point of alloys with some noble metals (such as Ag, Pd, Pt, and Rh), a pre-alloying strategy can be adopted through addition of other metals. Ag-Sb alloy with a eutectic point of 484 °C(*27*), Pd-Sn alloy with a eutectic point of 600 °C(*28*), Pt-Cd alloy with a eutectic point of 670 °C(*29*), and Rh-Sn alloy with a eutectic point of 660 °C(*30*), were used to enhance the atomic diffusion and intercalation of these noble metals in the gap between $Nb_2\square C$ layers. As a result, noble-metal-containing MAX phases, $Nb_2(Ag_{0.3}Sb_{0.4}Al_{0.3})C$ (figs. S28 and S29), $Nb_2(Pd_{0.5}Sn_{0.5})C$ (Figs. 2. E and F, and figs. S30 and S31), $Nb_2PtC$ (figs. S32 and S33), and $Nb_2(Rh_{0.2}Sn_{0.4}Al_{0.4})C$ (fig. S34), were synthesized. Interestingly, a disorder-to-order arrangement of guest atoms in interlayers was observed in MAX phases such as $Nb_2(Pt_{0.6}Al_{0.4})C$ (Fig. 3, and figs. S35 and S36) and $Nb_2(Au_{0.5}Al_{0.5})C$ (figs. S15). This phenomenon is attributed to the existence of vacancies $V_A$ between gaps of $Nb_2\square C$, which endows heavy atoms a high degree of freedom along the basal plane and low diffusion barrier. The ordered arrangement of guest atoms with host ones could efficiently alleviate the localized strain energy although the structure entropy is reduced. A similar order morphology was observed in intercalation of metal atoms into TMDs(*31*).

To apply the pre-alloying strategy to refractory metals, vanadium ($T_m\approx 1900$ °C) was first alloyed with zinc metal to form $V_4Zn_5$ with a liquidus temperature of 670 °C, and then reacted with LAMS-etched $Ti_3\square CN$. The simultaneous intercalation of V-Zn-Cu ternary alloys lead to the formation of $Ti_3(V_{0.4}Cu_{0.3}Zn_{0.3})CN$ MAX phase (Fig. 2G and fig. S37). Early transition metals are known to only occupy M-sites in MAX phases. The intercalation of vanadium into A layers implicates the possibility to tune the non-*vdW* bonding by *d*-block metals.

In addition to atomic replacement, a double layers of Bi atoms in a new MAX phase $Nb_2Bi_2C$ was observed (fig. S38), which means that this protocol not only allows the adjustment of the elemental composition but also the expansion of the structural diversity of the layered MAX-like phases, in analogy with the well-studied $Mo_2Ga_2C$(*32*).

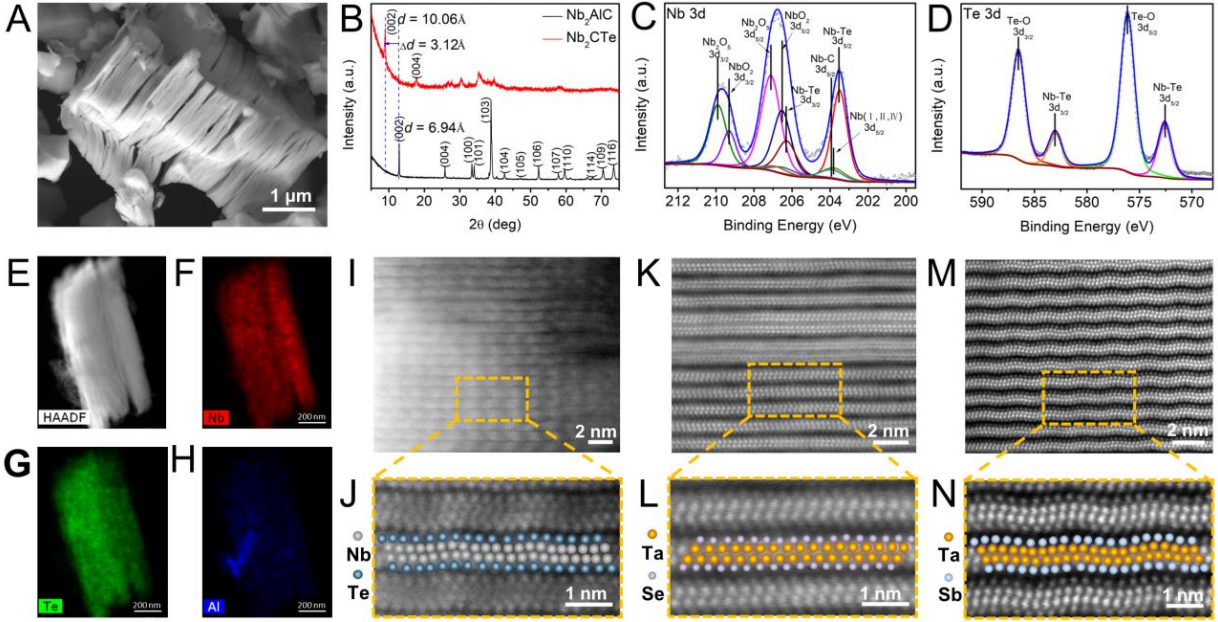

**Fig. 4. Transformation from MAX phase to MXene.** (**A**) SEM image of Nb$_2$CTe$_x$. (**B**) XRD patterns of Nb$_2$AlC and derived Nb$_2$CTe$_x$. (**C**) Nb 3$d$ and (**D**) Te 3$d$ XPS analysis of Nb$_2$CTe$_x$. (**E**) HAADF image of Nb$_2$CTe$_x$ and its corresponding EDS maps of Nb (**F**), Te (**G**), and Al (**H**). (**I**) STEM image of Nb$_2$CTe$_x$ along the [11$\bar{2}$0] direction and (**J**) its partial enlarged images with atomic structural model showing a ripple-like atomic structure. (**K**)-(**L**) STEM image of Ta$_2$CSe$_x$ and (**M**)-(**N**) Ta$_2$CSb$_x$.

## Derivation of MXenes from MAX phases

Halogenated MXenes (T = -Cl, -Br, and -I) were previously synthesized in corresponding halogenide molten salts(*14, 22*). The final termination of MXenes largely depends on the hard/soft acid/base (HSAB) principle if several ligands (Lewis bases) coexist in a molten salt. Most of transition metal cations (such as Ti$^{4+}$, Zr$^{4+}$, and V$^{5+}$) are typical hard Lewis acids with high positive charges (*33*). Consequently, the increase of chemical hardness in the halogen ligands (i.e., -I<-Br<-Cl<-F) strengthens the stability of resultant adducts, which explain the prevailing F-terminated MXenes through HF etching protocol(*11*). Accordingly, if a hard base anion S$^{2-}$ is added to the chloride melt, this chalcogenide ligand will exhibit a stronger coordination with the transition metal than Cl$^-$ which similar to the O$^{2-}$ prevail the F$^-$ terminal in HF-etched MXenes. In fact, S$^{2-}$ can be gradually released from FeS in molten salts and evidently coordinate with oxidized Ti in Ti$_2$□C to obtain S terminated MXene Ti$_2$CS$_x$ in a chloride melt (fig. S39).

Two more kinds of chalcogenide MXenes (T=-Te and -Se) were synthesized by the HSAB-guided ligand-selection coordination in molten chloride salts (figs. S40 to S44). The elemental Se ($T_m$≈220 °C, $r_{atom}$=117 pm) and Te ($T_m$≈450 °C, $r_{atom}$=143 pm) were directly employed with a LAMS scissor since they react with metal products to accelerate the kinetics of intercalation. For example, Cu$^{2+}$ LAMS scissor ($r_{ion}$=73 pm) first etched Al ($r_{atom}$=143 pm) out of Nb$_2$AlC and opened the gap between Nb$_2$C layers. Then the produced Cu ($r_{atom}$=128 pm) reacted with Te in the chloride melt to form Cu$_2$Te alloy with a eutectic point about 610 °C (figs. S45 and S46). The diluted Te atoms in Cu$_2$Te become active and withdraw electrons from electropositive Nb atoms to form Te$^{2-}$ anions, which further coordinate with oxidized Nb in Nb$_2$□C. An accordion-like morphology is shown in the SEM image of the resultant Nb$_2$CTe$_x$ MXene (Fig. 4A). The appearance of 000$l$ peaks at low Bragg angles and disappearing MAX phase diffraction peaks (Fig.

4B) confirm the full transformation of Nb$_2$AlC to Nb$_2$CTe$_x$. XPS analysis further corroborates the Nb-Te bond ($E_{\text{Nb 3d}}$ = 203.5 and 206.3 eV) on the surface of Nb$_2$CTe$_x$ MXene (Figs. 4, C and D, and figs. S47 and S48)(*34*). Both STEM-EDS and SEM-EDS analyses semi-quantitatively confirm the termination stoichiometry of $x \approx 1$ in Nb$_2$CTe$_x$ which clearly manifests a different coordination structure by two-electron chalcogen ligands when compared to the commonly inferred $x$ value of 2 in full-halogen-terminated MXenes, (Figs. 4, E to H, and fig. S49) (*14, 22, 24*). The Te terminations can be distinguished on both sides of Nb$_2$C layers along the zone axis [11$\bar{2}$0] in STEM image, and ripple-like atomic arrangement can be seen (Figs. 4, I and G, and fig. S50). This should be attributed to lattice stress caused by a large ionic radius of Te ligand ($r_{\text{ion}}$=221 pm). Actually, the lattice parameters ($a \approx 3.403$ Å, $c \approx 20.130$ Å) of Nb$_2$CTe$_x$ are significantly larger than that of the parent MAX phase Nb$_2$AlC ($a \approx 3.106$ Å, $c \approx 13.888$ Å) (fig. S51)(*35*). The enlarged $a$ value (about 9.5% increase) indicates an in-plane tensile stress exerted by Te ligand on the Nb$_2$C layers, which well explains the low-stoichiometry of ligand in Nb$_2$CTe$_x$ that efficiently reduces the strain energy. In fact, the ripple-like morphology almost disappears in MXene Ta$_2$CSe$_x$ (Figs. 4, K and L, and fig. S52) whose ligand Se has a small ion radius of 198 pm. It is noteworthy that Lewis bases -Se and -Te are softer than -S, and thus prefer to coordinate with soft acids, such as Nb and Ta cations, to form stable coordination structures in MXenes.

The same principle applies to surface modification of MXenes by phosphorus group elements. We observed, for example, that after etching of Ti$_3$AlC$_2$ by a LAMS scissor CuBr$_2$, P atoms in Cd$_3$P$_2$ with a low eutectic point of 740 °C subsequently intercalated the gap of Ti$_3$□C$_2$ and oxidized Ti atoms. As-formed P$^{3-}$ anions simultaneously became ligands with Br anions in the Ti$_3$C$_2$(P$_{0.4}$Br$_{0.6}$)$_x$ adduct (figs. S53 and S54). Antimony ($T_m \approx 630$ °C, $r_{\text{atom}}$=206 pm) also follows the same intercalation mechanism. Two antimony-terminated MXenes, Ta$_2$CSb$_x$ and Ta$_4$C$_3$Sb$_x$, were synthesized (figs. S55 to S58). The ripple-like morphology was also observed in MXene Ta$_2$CSb$_x$ since the atomic radius of Sb is similar to that of the neighboring Te (Figs. 4, M and N). It is worth noting that Sb can either occupy the A site as a metal atom in a MAX phase (**Route II**) or become a ligand as an anion in MXene (**Route IV**) due to its semi-metallic amphoteric character. Moreover, a series of MXenes with hybrid terminations are successfully synthesized (figs. S59 to S61).

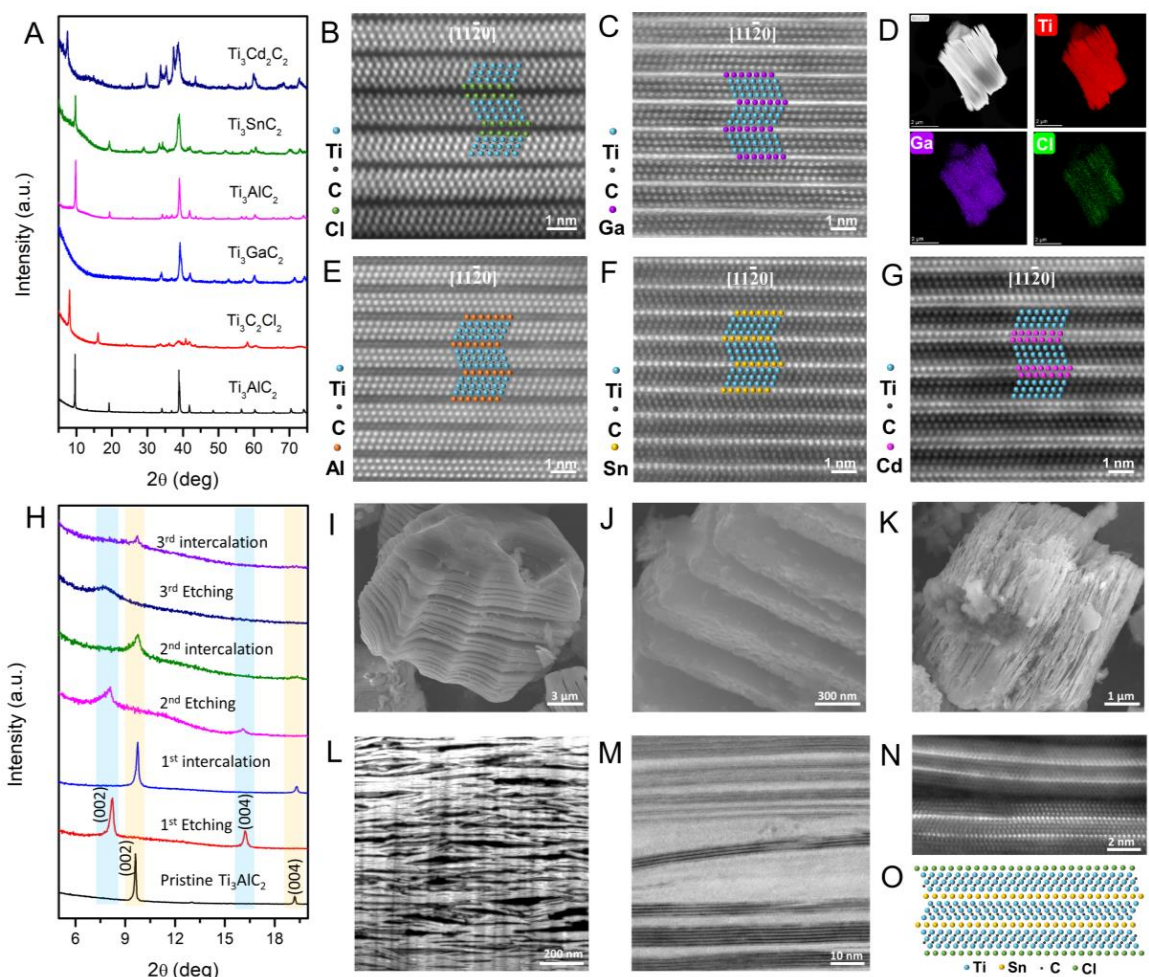

**Fig. 5. Reconstruction of MAX phases from MXenes.** (**A**) XRD patterns of the conversion of $Ti_3C_2Cl_2$ MXene to different MAX phases. STEM image of (**B**) $Ti_3C_2Cl_2$ MXene and (**C**) $Ti_3GaC_2$. (**D**) HADDF image of $Ti_3GaC_2$ and its corresponding EDS maps, showing the distribution of Ti, Ga, and Cl. STEM images of (**E**) $Ti_3AlC_2$, (**F**) $Ti_3SnC_2$, and (**G**) $Ti_3Cd_2C_2$. (**H**) XRD patterns showing the products of $CdCl_2$ LAMS etching and Al metal intercalation of $Ti_3AlC_2$ for three cycles. SEM images of $Ti_3AlC_2$ obtained from (**I**) the first, (**J**) second, and (**K**) third etching/intercalation cycle, respectively. (**L**) The cross-section image of the product after the third etching/intercalation, showing the stacked structure of the nanosheets. (**M**) Bright field STEM image of Sn-intercalated $Ti_3C_2$ at a low magnification. (**N**) Dark field STEM image of Sn atom-intercalated $Ti_3C_2$ at a high magnification and (**O**) its corresponding atomic models.

**Reconstruction of MAX phases from MXenes**.

The Lewis basic ligands on MXenes can be further removed by appropriate metal scissors, as demonstrated in a resultant intermediate $Ti_3\square C_2$ phase after -Cl ligands have been removed by the scissor metal, Al (fig. S62). Thus, it offers a route for interlayer structural editing, like for MAX phase reconstruction. Taking the well-known $Ti_3C_2Cl_2$ MXene as an example, the scissor metal Ga reacted with -Cl ligands to form gaseous $GaCl_3$ ($T_b$=201°C), which drove the transformation of $Ti_3C_2Cl_2$ into a non-terminated $Ti_3\square C_2$ (**Route III**). The resultant re-opened gap in $Ti_3\square C_2$ is available to either accommodate metal atom for MAX phase synthesis (**Route II**) or to coordinate other ligands for MXenes (**Route IV**). Consequently, the intercalation of excess Ga metal into as-formed gap of Ti$_3\square$C$_2$ recovers the MAX phase Ti$_3$GaC$_2$ as confirmed by atomically resolved STEM and EDS results which demonstrate the fully removal of -Cl terminations and occupancy of Ga atoms (Figs. 5, C and D). XRD results verify the phase conversion from the original Ti$_3$AlC$_2$ MAX phase via intermediate Ti$_3$C$_2$Cl$_2$ MXene to several MAX phases besides Ti$_3$GaC$_2$, such as Ti$_3$AlC$_2$ and Ti$_3$SnC$_2$ (Fig. 5A). Metals like Ga, Al, and Sn act as both metal scissors and intercalants (Figs. 5, E to G). Using Zn and Al as the scissors, a series of unconventional Cd-containing MAX and related nanolaminar phases, such as Ti$_3$(Cd$_{0.5}$Zn$_{0.5}$)C$_2$ and even double-A Ti$_3$Cd$_2$C$_2$, were reconstructed from Ti$_3$C$_2$Cl$_2$ MXene by intercalation of Cd atoms into defunctionalized Ti$_3\square$C$_2$, which are difficult to synthesize through either conventional sintering or isomorphous replacement reaction from bulk MAX phases. Most of reconstructed MAX phase particles preserve the accordion-like morphology of multilayer MXenes (fig. S63), indicating that such atomic reconstruction only happens in multilayer MXene lamellas stacked together by *vdW* force. This also suggests a possibility of synthesis of new materials with properties different from the currently known ones, including ferromagnetic and antiferromagnetic materials, catalysts, etc.

The double-closed-loop structure transitions between MAX phases and MXenes make protocol much flexible to edit their composition and structure (Fig. 1A). For example, the LAMS scissor-mediated etching (**Route I** and **IV**) of MXene-derived MAX phase through metal scissor-mediated reconstruction (**Route III** and **II**) leads to new MXene with a secondary accordion structure (Figs. 5, H to K). Meanwhile, the thickness of multilayer MXene lamellas in the secondary accordion structure is largely reduced (Fig. 5L). Moreover, this chemical-scissor-mediated structure editing protocol provides a strategy to directly achieve delamination of MXenes and MAX phases in molten salts. Delaminated Ti$_3$C$_2$Cl$_2$ MXene was synthesized through repeating interlayer editing alternately by CdCl$_2$ LAMS etching and Al metal intercalation of Ti$_3$AlC$_2$ (Fig. 5K). When metal scissor mediated atom intercalation is in the final step, it leads to discovery of a series of 2D carbides in which adjacent M$_{n+1}$X$_n$ slabs are intercalated by a layer of metal atoms but keep termination groups on the surface. Therefore, this metal-intercalated carbides combine the functional features of MXenes, such as tunable surface termination, and structural feature of MAX phases, such as oxidation-resistant interlayers. Atom-intercalated 2D carbides with the formula M$_{(n+1)m}$I$_{m-1}$X$_{nm}$T$_x$ (where I denotes the intercalated metal), can be determined if m layers of stacked MXenes M$_{(n+1)}$X$_n$T$_x$ were intercalated by m-1 layers of guest atoms (Fig. 5M). In order to promote the formation of 2D carbides in a delaminated few-layer Ti$_3$C$_2$Cl$_2$ MXene, a chemically inert element, Sn, was used as a metal scissor which drives out the -Cl ligands (**Route III**) that account for the interlayer *vdW* force in few-layer Ti$_3$C$_2$Cl$_2$ MXenes but leaves the surface terminations intact, and extra Sn finally intercalates into MXenes following **Route II** (fig. S64). A Sn-intercalated 2D carbide Ti$_6$SnC$_4$Cl$_2$ is thus obtained where n=2 and m=2. If m=1, there is no intercalation at all, and such atom-intercalated 2D carbide has the same formula as MXene (Figs. 5, N and O).

The chemical-scissor-mediated structural editing of MAX phases and their derived MXenes provides a broadly applicable protocol to engineer the structure and composition of both non-*vdW* and *vdW* layered materials. The regulated intercalation routes allow incorporation of unconventional elements into monoatomic layer of MAX phases which cannot be achieved through traditional metallurgic reaction. The noble metal atoms confined in monoatomic interlayers of MAX phases may produce highly active catalysts. Moreover, the atomic level stitching of MXenes into MAX phases by aid of metal scissors constitutes an unexplored route to create artificial 3D materials from 2D building blocks and offers a scalable process to delaminate MXenes directly in molten salts. Metal-intercalated 2D carbides, which combine the distinct features of MAX phases and MXenes, expand the range of known layered materials.